\documentstyle{aipproc}
\begin{document}
\raggedbottom
\widowpenalty=10000

\title{Off-Mass-Shell {\boldmath $\pi$}N Scattering and 
{\boldmath $pp \rightarrow pp \pi^0$}  }

\author{M. T. Pe\~{n}a$^{*,**}$, S.\ A.\ Coon$^{\dagger}$, 
  J. Adam Jr.$^{\ddag}$, and A. Stadler$^{*,\star}$   }

\address
{$^{*}$ Centro de F\'{\i}sica Nuclear, 1699 Lisboa, Portugal\\
$^{**}$ CFIF,Instituto Superior T\'ecnico, 1096 Lisboa, Portugal\\
$^{\dagger}$ Physics Department,
        New Mexico State University, Las Cruces, NM  88003, USA\\
$^{\ddag}$Institute of Nuclear Physics, \v{R}ez n. Prague, CZ-25068, 
Czech Republic\\
$^{\star}$ Departamento de F\'{\i}sica, Universidade de \'Evora, 
7000 \'Evora, Portugal}

\maketitle   
\vspace{-.1in}

\begin{abstract}
	We adapt the off-shell $\pi$N amplitude of the Tucson-Melbourne
three-body force to  the half-off-shell amplitude of the pion
rescattering  contribution to  $pp \rightarrow pp \pi^0$ near
threshold.  This {\em pion} rescattering contribution, together with
the impulse term, provides a good description of the data when the 
half-off-shell amplitude is linked to the phenomenological  invariant
amplitudes obtained from meson factory $\pi$N scattering data.
\end{abstract}

\vspace{-.1in}

The precise measurements of $p p \rightarrow p p
\pi^0$~\cite{IUCF,Celsius} could be used to calibrate or constrain the 
$\pi$N scattering amplitude $F^+$ underlying 2$\pi$ exchange
three-nucleon forces,  in a manner  complementary to the standard
constraints of  on-mass-shell $\pi$N data  and the  implementation of
chiral symmetry~\cite{TM79,TM93}. To see this, consider a 
two-pion-exchange three-body-force diagram and strip off one  of the
outer nucleons so that  the emerging pion is on its mass shell.   The
result is the pion ``rescattering" diagram   found to be tiny if
assumed to be  proportional to the tiny isospin even s-wave $\pi$N
scattering length. However, the   off-mass-shell $\pi$N amplitudes of
PCAC-current algebra\cite{TM79,TM93}  have s-wave terms which are of
the same magnitude  as the p-wave terms familiar from $\Delta$-isobar
models. A qualitative estimate of $p p \rightarrow p p \pi^0$ due to
the impulse diagram plus half-off-mass-shell pion rescattering diagram
was given a long time ago by Hachenberg and Pirner~\cite{HP78}.  Later
calculations with half-off-shell amplitudes appear to  confirm the
Hachenberg-Pirner findings of an enhancement of the  cross section  via
$s$-wave pion rescattering~\cite{Efro,Oset}. In contrast, the pion
rescattering diagram calculated with  chiral perturbation theory
appears to decrease the theoretical cross section rather far below the
data~\cite{CHPT}.

We calculate in momentum space the non-relativistic impulse term plus
half-off-shell  pion rescattering term.  The $T$-matrix which enters
into the latter is
\begin{equation}
	T^{TM}_{\pi} = 
\frac{-i}{(2\pi)^3} \frac {g}{2m}
\mbox{\boldmath $\sigma$}_2\!\cdot\!\mbox{\boldmath $k$}	
\frac{1}{\mu^2 - k^2} (- \bar F^{+} - \Delta F^+)  \label{eq:T}
\end{equation}
where $k$ is the four-momentum of the pion exchanged between protons 1
and 2 ($\mbox{\boldmath $k$} = \mbox{\boldmath $p$}_2' - 
\mbox{\boldmath $p$}_2$), and $F$ represents the 
appropriate invariant amplitude of 
$ \pi(k) + N(p_1) \rightarrow \pi(q) + N(p'_1)$, (proton 1 emitting
the real pion).  The
Tucson-Melbourne (TM)  Z-graph
contribution (labeled $\Delta F^+$) is given in Refs. \cite{TM79,TM93} and
the (covariant nucleon pole removed)  non-spin flip even current algebra $\pi$N
amplitude for general pion four momenta $q,k$ is
\begin{equation}
  \bar F^{+}(\nu ,t,q^{2},k^{2}) = [(1-\beta)(\frac{q^{2}+k^{2}}{{\mu}^{2}}-1)+
\beta (\frac{t}{{\mu}^{2}}-1)] \frac{\sigma}
    {{f_{\pi}}^2}
      + C^{+}(\nu ,t,q^{2},k^{2})
 \label{eq:fampli}
\end{equation}
where $\sigma$ is the pion-nucleon $\sigma$ term,
$f_{\pi} \approx 93$ MeV, and $C^{+}$
contains the higher order $\Delta$ isobar contribution calculated
dispersively~\cite{ST}. The latter amplitude
must have the simple form \cite{TM79,ST}
\begin{equation}
   C^{+}(\nu ,t,q^{2},k^{2})=c_{1}{{\nu}^{2}}+c_{2} q \cdot k +O(q^{4})
\, .
\label{eq:cexpa}
\end{equation}
On the other hand, the assumed form of the multiplier of
$\sigma/f_{\pi}^2$ 
(adapted \cite{TM79,Praguelec} for $\pi$N scattering from the $SU(3)$
generalization of the Weinberg low energy expansion for $\pi\pi$
scattering)
is such that $\bar F^{+}$  satisfies the soft pion theorems.
The $c_2$ and $\beta$ constants in
the coefficient of the $q \cdot k$
term can be eliminated in favor
of the on-shell (measurable) quantity $\bar F^{+}(0,\mu^2,\mu^2,
\mu^2)$~\cite{TM79,TM93}.  We expand $\bar F^{+}$in powers of $q,k$ and 
drop terms
of ${\cal O}(\mu^2/m^2)$ to get a nonrelativistic amplitude with which we do
quantum mechanics. 
In the kinematics of the Tucson-Melbourne $2\pi$ exchange three-body
force, the quantity $\nu^2$ is of ${\cal O}(\mu^4/m^2)$ and $c_2\nu^2$ is
therefore dropped from the (two pions off-mass-shell) amplitude.  It is
easy to see, however,  that, exactly at pion production threshold, the
needed values in $\bar F^+$ of (\ref{eq:T}) are $  \nu = \mu$,  $t=-m\mu$, 
$q^2 = \mu^2$,  $k^2= -m\mu$, so the quantity $c_{1}{{\nu}^{2}}$ must
be retained in a realistic calculation (which, by the way, should {\em
not} ``freeze" the amplitude at the threshold values).  The retention of
$c_{1}{{\nu}^{2}}$ and placing of $q^2$ on-shell for the  produced pion
are the only changes from the structure of the TM amplitude in the
three-nucleon force. We
follow Ref. \cite{FHvK} and remove a spurious  term from $(- \bar F^{+}
- \Delta F^+)$ in Eq. \ref{eq:T}; it corresponds to a pion produced
directly from a four-fermion (contact) interaction and should not be
present in a rescattering diagram.

 The three parameters, $\sigma/f_{\pi}^2\approx 1.35 \mu^{-1}$,  $\bar
F^{+}(0,\mu^2,\mu^2,\mu^2)\approx -0.08\mu^{-1} $, and and
$c_{1}\approx +1.23 \mu^{-3}$ of the present TM $\pi$N amplitude are found
from a recent interior dispersion analysis~\cite{Kaufmann}  of the SP98
phase shift solution to meson factory data.  The monopole $\pi^0$NN
vertex function of the exchanged pion reproduces the 2\%
Goldberger-Treiman discrepancy~\cite{Praguelec} suggested by the
current $\pi$N and NN data.  Our results are shown in Figure 1 and
compared with the data points  labeled IUCF and Celsius  from 
Refs.~\cite{IUCF,Celsius} respectively and with our calculation of the
ChPT~\cite{CHPT} treatment of pion rescattering.

\pagebreak

  \setcounter{figure}{0}
\begin{figure}[htpb]
\vspace{1.0in}
\unitlength1.cm
\begin{picture}(5,7)(0,0)
\includegraphics{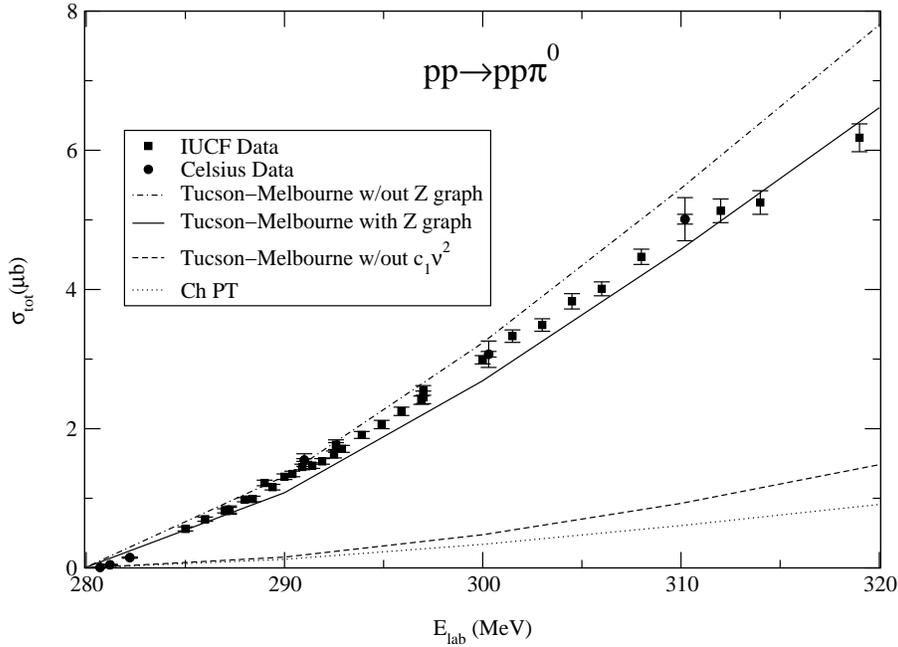}
\end{picture}
\caption{Cross section for  $p p \rightarrow p p \pi^0$ using the Bonn-B
 $NN$ potential for the initial and final state interaction of the two
protons. All calculations include both impulse and pion rescattering
diagrams. The ``frozen kinematics" approximation is not used.  }
\end{figure}

\end{document}